\documentclass[prc,superscriptaddress,unsortedaddress,twocolumn]{revtex4}
\usepackage[english]{babel}
\usepackage{graphicx,epsf,bm,amsmath,color}
\usepackage{here,ulem}
\def\bk{\mbox{\boldmath $k$}}
\def\bp{\mbox{\boldmath $p$}}
\def\bq{\mbox{\boldmath $q$}}
\def\bfsigma{\mbox{\boldmath $\sigma$}}

\def\bftau{\mbox{\boldmath $\tau$}}
\def\cg0#1#2#3{( #1 0 #2 0 | #3 0 )}
\def\sixj#1#2#3#4#5#6{\begin{Bmatrix}
                    #1 & #2 & #3 \\ #4 & #5 & #6 \end{Bmatrix} }
\def\h3t{\mbox{$_\Lambda^3$H}}
\begin{document}
\title{Contributions of $2\pi$-exchange, $1\pi$-exchange, and contact three-body
forces in NNLO ChEFT to \h3t}
\author{M. Kohno}
\affiliation{Research Center for Nuclear Physics, Osaka University, Ibaraki 567-0047,
Japan}

\author{H. Kamada}
\affiliation{Department of Physics, Faculty of Engineering, Kyushu Institute of Technology,
Kitakyushu 804-8550, Japan}

\author{K. Miyagawa}
\affiliation{Research Center for Nuclear Physics, Osaka University, Ibaraki 567-0047,
Japan}
\begin{abstract}
Faddeev calculations of hypertriton (\h3t) separation energy are performed,
incorporating all next-to-next-to-leading-order $\Lambda$NN three-body
forces (3BFs) in chiral effective
field theory: $2\pi$-exchange, $1\pi$-exchange, and contact interactions.
The $1\pi$-exchange and contact interactions are rewritten in a form
suitable for evaluating partial-wave matrix elements. The $\Lambda$-deuteron
folding potentials constructed from these 3BFs are evaluated
to demonstrate their contributions to \h3t. The $1\pi$-exchange
interaction provides an attractive effect in which the $d$-state component
of the deuteron wave function plays an important role. The attractive contribution
tends to cancel the repulsive ones from the $2\pi$-exchange and contact 3BFs.
Faddeev calculations show that the net effect of the 3BFs to the \h3t separation
energy is small in a range between $-5$ to $+20$ keV, depending on the NN
interaction used. Although these results are based on speculative
low-energy constants, they can serve as a reference for further investigations.
\end{abstract}

\maketitle
\section{Introduction}
The hypertriton (\h3t) is an isospin $T=0$ $\Lambda$NN bound state with small
separation energy to $\Lambda$ + deuteron. Because a $\Lambda$N two-body
bound system is unlikely, \h3t is valuable to investigate $\Lambda$-nucleon
interaction. For example, its binding energy helps resolve the spin singlet and
triplet strengths of the $\Lambda$N interaction to compensate for the lack of
scattering data. Although the separation energy has not been accurately established,
the situation will change shortly due to ongoing new experiments.
On the theory side, the studies of the $\Lambda$NN system bear some ambiguities
from the effects of possible three-body forces (3BFs). In ordinary light nuclei,
an attractive contribution of three-nucleon forces is necessary to explain their
binding energies accurately. Likely, we may expect a non-negligible effect of the
3BFs in \h3t. It is needed to study whether this contribution is attractive or
repulsive and the order of magnitude.

We have studied the effect of the 3BF in chiral effective field theory (ChEFT) for the
hypertriton in Ref. \cite{KKM23}. In that article, the $2\pi$-exchange $\Lambda$NN
interaction derived by Petschauer \textit{et~al.} \cite{PET16} at the next-to-next-to-leading-
order (NNLO) was addressed as an initial attempt. Faddeev calculations were carried
out for the first time, incorporating the 3BF matrix-elements, to show that the contribution is
repulsive of the order of 20 keV. This size is small but not negligible compared with the
small separation energy of the hypertriton, the present world average of which is
$148\pm40$ keV \cite{MA22}.

The remaining NNLO $\Lambda$NN 3BFs, a $1\pi$-exchange 3BF and a contact
3BF, are considered in this article. Because the expressions of these 3BFs in momentum space
presented in Ref. \cite{PET16} are not readily applicable in evaluating partial-wave matrix elements,
we rewrite them in a form appropriate for use. As for low-energy coupling constants (LECs),
we rely on the estimation by Petschauer \textit{et~al.} \cite{PET17} in a decouplet-dominant
model. Although there are uncertainties, it is worth to evaluate the effect of these 3BFs
by using those LECs as reference values for further studies.

Expressions of the $1\pi$-exchange $\Lambda$NN 3BF and the contact $\Lambda$NN
term are rewritten in Sec. 2 and Sec. 3, respectively, to be suitable for evaluating partial-wave
matrix elements. Before explicit Faddeev calculations are carried out for \h3t, it is worth
obtaining a preliminary idea about the contributions of the 3BFs by evaluating
a $\Lambda$-deuteron folding potential as was done for the $2\pi$-exchange 3BF
in Ref. \cite{KKM23}. Those $\Lambda$-deuteron folding potentials in momentum
space are discussed in Sec. 4. Results of Faddeev calculations of L3H,
which include these 3BFs along with the 2pi-exchange 3BF, are presented in Sec 5.
Summary follows in Sec 6. Summary follows in Sec. 6.

\section{one-pion exchange $\Lambda$NN}
\begin{figure}[bht]
\begin{center}
\includegraphics[width=0.35\textwidth]{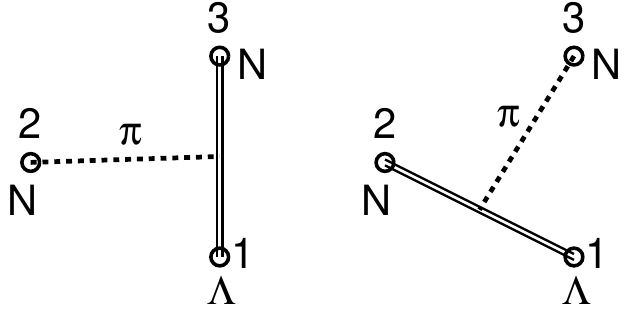}
\end{center}
\caption{$1\pi$-exchange $\Lambda NN$ 3BF corresponding to Eq. \ref{eq:1pi}.
The double line means a $\Lambda$N contact part smeared by a cutoff
factor.}
\label{fig:lnct}
\end{figure}

The $1\pi$-exchange $\Lambda$NN interaction in momentum space in the NNLO was
shown by Petschauer \textit{et~al.}  in Ref. \cite{PET16} as follows:
\begin{align}
 &V_{1\pi}^{\Lambda NN} =-\frac{g_A}{2f_0^2} (\bftau_2\cdot\bftau_3) \nonumber \\
 & \times \left\{ \frac{\bfsigma_2\cdot\bq_{2'2}}{\bq_{2'2}^2+m_\pi^2}
  (D_1'\bfsigma_1+D_2'\bfsigma_3)\cdot \bq_{2'2}  \right. \nonumber \\
 & \hspace{3mm} +\frac{\bfsigma_3\cdot\bq_{3'3}}{\bq_{3'3}^2+m_\pi^2}
  (D_1'\bfsigma_1+D_2'\bfsigma_2)\cdot \bq_{3'3} \nonumber \\
 & +P_{23}^{(\sigma)}P_{23}^{(\tau)}P_{13}^{(\sigma)}
 \frac{\bfsigma_2\cdot\bq_{3'2}}{\bq_{3'2}^2+m_\pi^2}
 \left[-\frac{D_1'+D_2'}{2}(\bfsigma_1+\bfsigma_3)\cdot\bq_{3'2}   \right. \nonumber \\
 &\left.\hspace{2cm}+\frac{D_1'-D_2'}{2}i(\bfsigma_3\times\bfsigma_1)\cdot\bq_{3'2}\right]
 \nonumber \\
 & +P_{23}^{(\sigma)}P_{23}^{(\tau)}P_{12}^{(\sigma)}
\frac{\bfsigma_2\cdot\bq_{2'3}}{\bq_{2'3}^2+m_\pi^2}
 \left[-\frac{D_1'+D_2'}{2}(\bfsigma_1+\bfsigma_2)\cdot\bq_{2'3} \right. \nonumber \\
 &\left. \left. \hspace{2cm}- \frac{D_1'-D_2'}{2}i(\bfsigma_1\times\bfsigma_2)
 \cdot\bq_{2'3}\right]\right\},
\end{align}
in which a label of 1 assigned to the $\Lambda$ hyperon.
$g_A$ is the axial-vector coupling constant, $f_0$ is the pion decay constant, $\bq_{i'j}$ is
a momentum transfer at each $\pi$NN vertex, and $P_{ij}^{(\sigma)}$ ($P_{ij}^{(\tau)}$)
is an exchange operator in spin (isospin) space.
Explicit numbers of the low-energy constants (LECs), $D_1'$ and $D_2'$, are set in Se. IV. 
Noting $P_{ij}^\sigma =\frac{1}{2}(1+\bfsigma_i\cdot\bfsigma_j)$ and
 $\sigma_k \sigma_\ell=\delta_{k\ell}+i\epsilon_{k\ell m}\sigma_m$, the above
expression is rewritten as
\begin{align}
 V_{1\pi}^{\Lambda NN} =& -\frac{g_A}{2f_0^2}\left(1-P_{23}^{(\sigma)}
 P_{23}^{(\tau)}P_{23}^{(p)}\right)(\bftau_2\cdot\bftau_3) \nonumber \\
 & \times \left\{ \frac{\bfsigma_2\cdot\bq_{2'2}}{\bq_{2'2}^2+m_\pi^2}
  (D_1'\bfsigma_1+D_2'\bfsigma_3)\cdot \bq_{2'2} \right.  \nonumber \\
 & \left. \hspace{1em}+\frac{\bfsigma_3\cdot\bq_{3'3}}{\bq_{3'3}^2+m_\pi^2}
  (D_1'\bfsigma_1+D_2'\bfsigma_2)\cdot \bq_{3'3}  \right\},
\end{align}
where $P_{23}^{(p)}$ stands for the exchange operator for the momenta of the nucleon pair,
and $P_{23}^{(\sigma)}P_{23}^{(\tau)}P_{23}^{(p)}$ means an exchange operator for the
two nucleons. The factor $1-P_{23}^{(\sigma)}P_{23}^{(\tau)}P_{23}^{(p)}$ can be taken
care of by the anti-symmetrization of the pair of two-nucleon wave functions. In that case,
$P_{23}^{(\sigma)}P_{23}^{(\tau)}P_{23}^{(p)}$ is unnecessary, and
the $1\pi$-exchange $\Lambda$NN interaction should be considered as follows.
\begin{align}
 V_{1\pi}^{\Lambda NN} =& -\frac{g_A}{2f_0^2} (\bftau_2\cdot\bftau_3) \nonumber \\
 & \times \left\{ \frac{\bfsigma_2\cdot\bq_{2'2}}{\bq_{2'2}^2+m_\pi^2}
  (D_1'\bfsigma_1+D_2'\bfsigma_3)\cdot \bq_{2'2} \right. \nonumber \\
 & \left. \hspace{1em}+\frac{\bfsigma_3\cdot\bq_{3'3}}{\bq_{3'3}^2+m_\pi^2}
  (D_1'\bfsigma_1+D_2'\bfsigma_2)\cdot \bq_{3'3}  \right\}.
\label{eq:1pi}
\end{align}
The first (second) term in the curly brackets of the above equation corresponds to the
left (right) diagram in Fig. \ref{fig:lnct}.

Now, this expression is transformed to the partial-wave expansion form applicable
in evaluating matrix elements using the expression given in Ref. \cite{KKM22}.
\begin{align}
 & V_{1\pi}^{\sigma_2} = 4\pi (\bftau_2\cdot\bftau_3)\sum_{K=0,1,2} \sum_{\ell_a\ell_b}
  \nonumber \\
  & \left\{ V_{1\pi,(2,3)}^{K,\ell_a,\ell_b}(p,q) [[\bfsigma_2\times\bfsigma_3]^K
 \times [Y_{\ell_a}(\hat{\bp})\times Y_{\ell_b}(\hat{\bq})]^K]^0\right. \nonumber \\
 & +V_{1\pi,(1,2)}^{K,\ell_a,\ell_b}(p,q) [[\bfsigma_1\times\bfsigma_2]^K
\times [Y_{\ell_a}(\hat{\bp})\times Y_{\ell_b}(\hat{\bq})]^K]^0 \nonumber \\
 & +\left.V_{1\pi,(3,1)}^{K,\ell_a,\ell_b}(p,q)
 [[\bfsigma_3\times\bfsigma_1]^K\times [Y_{\ell_a}(\hat{\bp})
 \times Y_{\ell_b}(\hat{\bq})]^K]^0 \right\},
\label{eq:1pipw}
\end{align}
where an abbreviated notation for a tensor product with Clebsch-Gordan coefficients is used:
\begin{align}
 [T_{j_1}\times T_{j_2}]_M^J =\sum_{m_1 m_2} (j_1 m_1 j_2 m_2|JM)T_{j_1 m_1}T_{j_2 m_2}.
\end{align}
$\bp$ and $\bq$ are differences of the final and initial Jacobi momenta. That is, denoting
each momentum of the $i$-th initial particle by $\bk_i$, Jacobi momenta are defined as
$\bp_1=\bk_2-\bk_3$ and $\bq_1= \bk_1$ in the center-of-mass frame. Jacobi momenta
for the final configuration are represented with a prime. Then, $\bp\equiv \bp_1'-\bp_1$
and $\bq'\equiv \bq_1'-\bq_1$. 
The explicit expression of  $V_{1\pi,(i,j)}^{K,\ell_a,\ell_b}(p,q)$ is given in Appendix A.

\section{contact term}
The contact $\Lambda$NN term in the NNLO was shown by Petschauer \textit{et~al.}
in Ref. \cite{PET16} as follows:
\begin{align}
 V_{ct}^{\Lambda NN} =& C_1'(1-\bfsigma_2\cdot\bfsigma_3)(3+\bftau_2\cdot\bftau_3)
 \nonumber \\
  & +C_2'(\bfsigma_\Lambda \cdot(\bfsigma_2+\bfsigma_3))(1-\bftau_2\cdot\bftau_3)
 \nonumber \\
 & +C_3'(3+\bfsigma_2\cdot\bfsigma_3)(1-\bftau_2\cdot\bftau_3),
\end{align}
in which a label of 1 is assigned to the $\Lambda$ hyperon. The exchange of the
nucleon pair is explicitly taken care of, and the expression is
antisymmetric under the exchange of two nucleons; that is,
$P_{23}^{(\sigma)}P_{23}^{(\tau)}P_{23}^{(p)}V_{ct}^{\Lambda NN}=-V_{ct}^{\Lambda NN}$.
Therefore, $V_{ct}^{\Lambda NN}$ is written as
$\frac{1}{2}(1-P_{23}^{(\sigma)}P_{23}^{(\tau)}P_{23}^{(p)})V_{ct}^{\Lambda NN}$.
It means that when the interaction is applied to the $\Lambda$NN wave function in which
two nucleons are antisymmetrized, factor $\frac{1}{2}$ is necessary.


The above expression is rewritten in the following partial-wave expansion form.
\begin{align}
  V_{ct}^{\Lambda NN} =4\pi \sum_{\tau=0,1} \sum_{k,m,n} V_{ct}^{(k,m,n,\tau)}
  (\bftau_2\cdot\bftau_3)^\tau \nonumber \\
  \times [[\bfsigma_\Lambda^{(k)} \times [\bfsigma_2^{(m)}\times \bfsigma_3^{(n)}]^k]^0
 \times [Y_0(\hat{\bp})\times Y_0(\hat{\bq})]^0]^0.
\label{eq:ct}
\end{align}
The coefficients $V_{ct}^{(k,m,n,\tau)}$ are given in Table I.

\begin{table}[b]
\begin{center}
\begin{ruledtabular}
\begin{tabular}{ll}
 $V_{ct}^{(0,0,0,0)}=\frac{1}{4\pi}\frac{3}{2}(C_1'+C_3')$ & $V_{ct}^{(0,0,0,1)}
 =\frac{1}{4\pi}\frac{1}{2}(C_1'-3C_3')$\\
 $V_{ct}^{(0,1,1,0)}=\frac{1}{4\pi}\frac{\sqrt{3}}{2}(3C_1'-C_3')$ &
 $V_{ct}^{(0,1,1,1)}=\frac{1}{4\pi}\frac{\sqrt{3}}{2}(C_1'+C_3')$ \\
 $V_{ct}^{(1,1,0,0)}=-\frac{1}{4\pi}\frac{\sqrt{3}}{2}C_2'$ & $ V_{ct}^{(1,1,0,1)}
 =\frac{1}{4\pi}\frac{\sqrt{3}}{2}C_2'$ \\
  $V_{ct}^{(1,0,1,0)}=-\frac{1}{4\pi}\frac{\sqrt{3}}{2}C_2'$ & $V_{ct}^{(1,0,1,1)}
 =\frac{1}{4\pi}\frac{\sqrt{3}}{2}C_2'$ \\
\end{tabular}
\caption{Coefficients $V_{ct}^{(k,m,n,\tau)}$ of the NLO contact $\Lambda$NN
interaction written in the form of Eq. (\ref{eq:ct}). The factor of $\frac{1}{2}$ explained
in the text is included. Other coefficients not shown are zero. 
}
\end{ruledtabular}
\end{center}
\end{table}

\section{$\Lambda$-deuteron folding potential}
It is instructive to evaluate the $\Lambda$-deuteron folding potential provided
by the $1\pi$-exchange and the contact 3BFs to demonstrate the
contribution of these 3BFs in the hypertriton.
The folding potential is calculated by the following integration:
\begin{align}
 U_{\Lambda-d}^{J_t}(q_1',q_1)=& \iint p_1'^2 dp_1' p_1^2 dp_1
 \langle [\Psi_d(\bp_1'),(\ell_{\Lambda}' 1/2)j_\Lambda]J_t| \notag \\
  & \times V_{1\pi (ct)}^{\Lambda NN}| [\Psi_d(\bp_1),(\ell_{\Lambda} 1/2)j_\Lambda]J_t
   \rangle, \label{eq:lpd} \\
 \Psi_d(\bp_1)=&\sum_{\ell_d=0,2} \frac{1}{p_1}\phi_{\ell_d}(p_1)[Y_{\ell_d}(\hat{\bp_1})
 \times \chi_d^{1}]_{m}^1,
\label{eq:sd}
\end{align}
where $\Psi_d(\bp)$ represents a deuteron wave function. A detailed calculational
procedure for the 3BF in the form of Eq. \ref{eq:1pi} or \ref{eq:ct} is given in Appendix B
in Ref. \cite{KKM22}.

Figure \ref{fig:1pi} shows the result of the $1\pi$-exchange $\Lambda$NN 3BF
given in Eq. (\ref{eq:1pi}) with the total angular momentum $J_t=1/2$.
The LECs are taken from the estimation by Petschauer \textit{et~al.} \cite{PET17}:
$D_1'=0$ and $D_2'=\frac{2CH'}{9\Delta}$ with $C\approx\frac{3}{4}g_A$,
$H'\approx 1/f_0^2$, and the decuplet-octet baryon mass splitting $\Delta$.
The numerical value of  $D_2'$ is set as $D_2'=3.268\times 10^{-3}$ fm$^2$MeV$^{-1}$
using $g_A=1.29$, $f_0=92.4$ MeV, and $\Delta=300$ MeV.
As discussed in Ref. \cite{PET17}, the sign of $D_2'$ could be the opposite.
The upper panel of Fig. \ref{fig:1pi} depicts the contribution of the $s$-wave
pair of the bra and ket deuteron wave functions. The repulsive magnitude is
smaller than the corresponding strength of the $2\pi$-exchange
$\Lambda$NN reported in Ref. \cite{KKM23}. The result of the $s$-$d$ pair of
the bra and ket deuteron wave functions is shown in the lower panel of Fig. \ref{fig:1pi},
which is attractive, and the magnitude is unexpectedly large despite the small $d$-wave
component of the deuteron wave function. The $d$-$s$ pair provided the same contribution.
The contribution of the $d$-wave pair of the bra and ket deuteron wave functions is
negligible. The net $\Lambda$-deuteron folding potential, including the contribution of
the $2\pi$-exchange reported in Ref. \cite{KKM23}, is found to be attractive and to have
a value of about $-0.35$ MeV at the origin.

\begin{figure}[thb]
\includegraphics[width=0.4\textwidth]{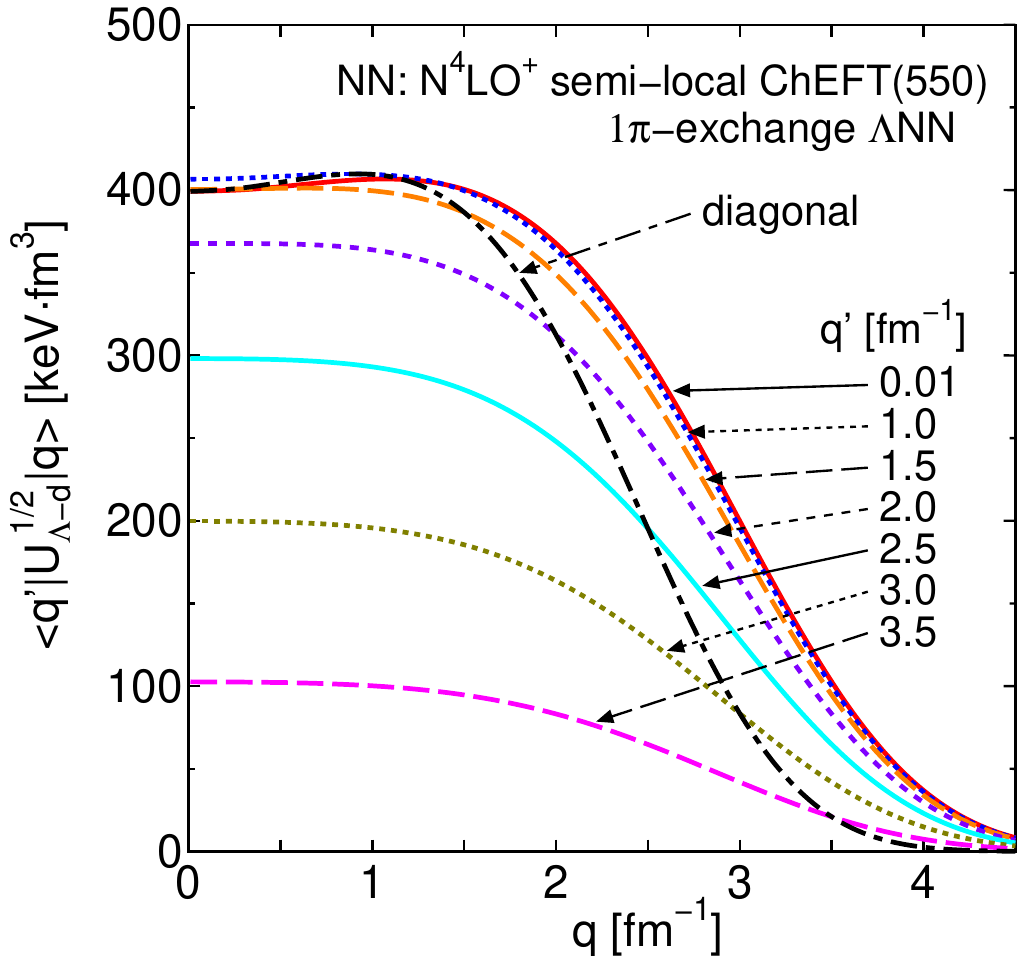}
\includegraphics[width=0.4\textwidth]{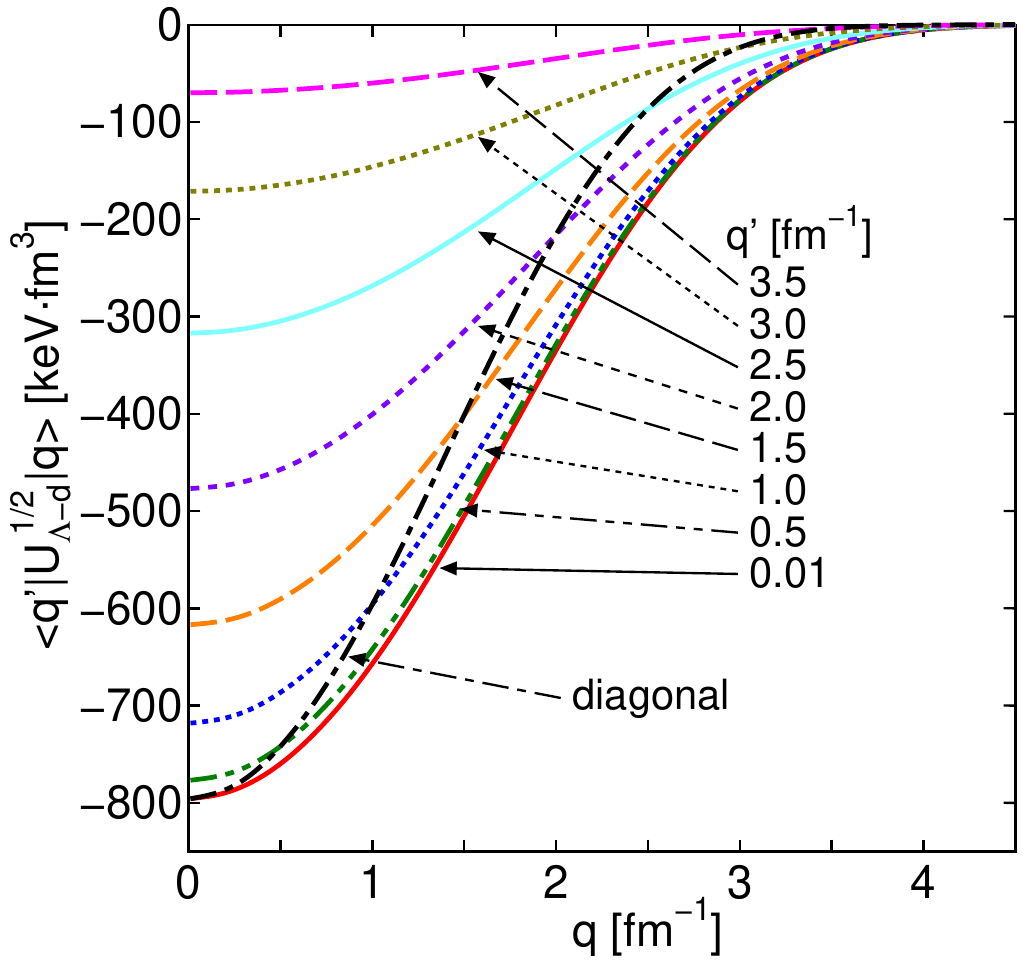}
\caption{$\Lambda$-deuteron folding potential provided by the $1\pi$-exchange
$\Lambda$NN 3BF given in Eq. (\ref{eq:1pi}). The coupling constants employed
are $D_1'=0$ and $D_2'=3.268\times 10^{-3}$ fm$^2$MeV$^{-1}$ based on the
estimation by Petschauer \textit{et~al.} in Ref. \cite{PET17} using a decouplet
saturation model. The deuteron wave functions are calculated using
the N$^4$LO$^+$ NN interactions \cite{RKE18} in chiral effective field theory.
}
\label{fig:1pi}
\end{figure}

\begin{figure}[thb]
\includegraphics[width=0.4\textwidth]{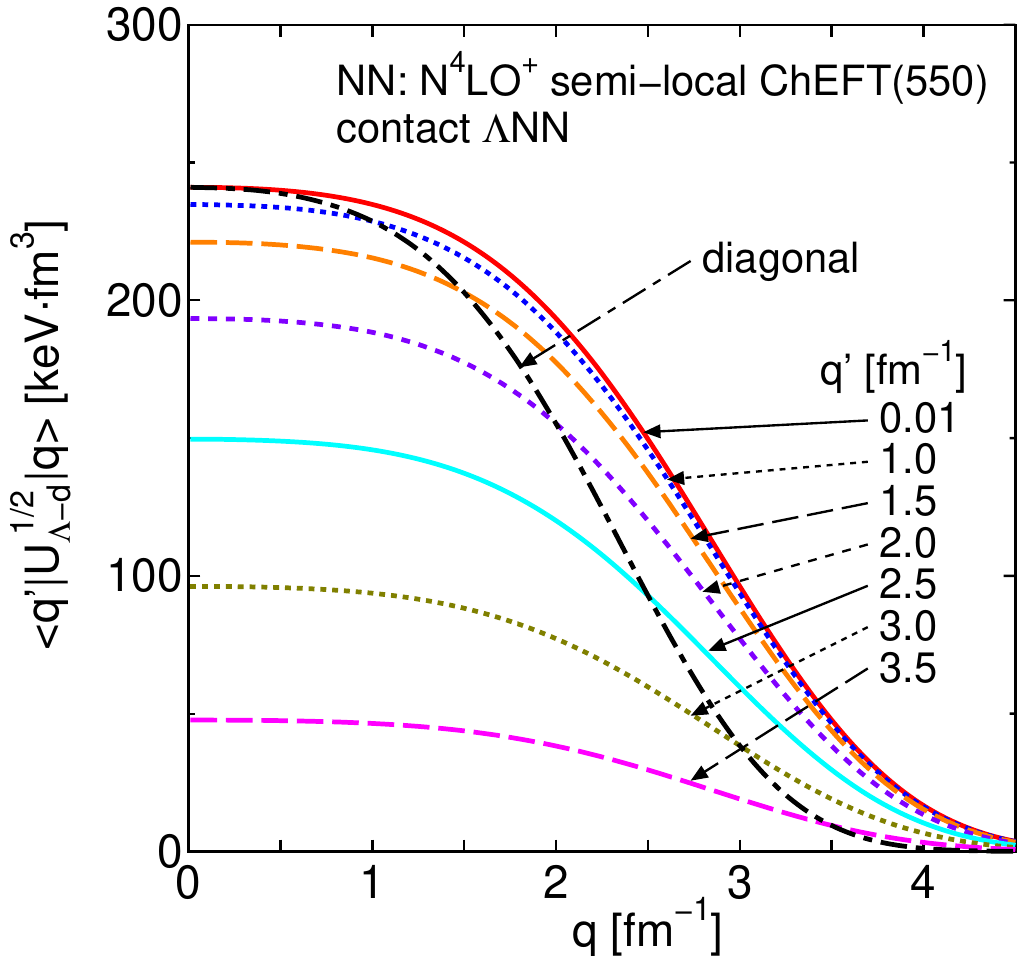}
\caption{$\Lambda$-deuteron folding potential provided by the
contact $\Lambda$NN 3BF given in Eq. (\ref{eq:ct}). The coupling constants
employed are $C_1'=C_3'=\frac{1}{f_0^4 \Delta}$ and $C_2'=0$ with $f_0=92.4$ MeV
and $\Delta=300$ MeV,
which are those estimated by Petschauer \textit{et~al.} in Ref. \cite{PET17}
using a decouplet saturation model.
}
\label{fig:ct}
\end{figure}

Figure \ref{fig:ct} shows the result of the contact term of the $\Lambda$NN 3BF
given in Eq. (\ref{eq:ct}). The low-energy constants are set as
$C_1'=C_3'=\frac{1}{f_0^4 \Delta}$ and $C_2'=0$ with $f_0=92.4$ MeV and
$\Delta=300$ MeV, following the estimation by Petschauer \textit{et~al.} \cite{PET17}.
Other contributions from the $s$-$d$, $d$-$s$, and $d$-$d$
pairs of the bra and ket deuteron wave functions are negligibly small.

Therefore, the sum of the contributions of the $1\pi$-exchange and contact $\Lambda$NN
3BFs is attractive. Even if the repulsive contribution of the $2\pi$-exchange reported
in Ref. \cite{KKM23} is included, the net contribution of the three types of the NLO
$\Lambda$NN 3BFs with the LECs estimated by Petschauer \textit{et~al.} \cite{PET17}
is expected to be attractive at around $-300$ keV.

\section{Faddeev calculations of $_\Lambda^3$H}
In this section, we present the results of the explicit Faddeev calculations of the
separation energy of  $_\Lambda^3$H that include all NNLO $\Lambda$NN
interactions in chiral effective field theory.
The method of calculation is explained in Ref. \cite{KKM23}.
Although the LECs employed are speculative, the calculated results serve as
a reference number for possible 3BF contributions in $_\Lambda^3$H.

Uncertainties in the prediction of the \h3t separation energy due to the NN interaction
employed were discussed in Ref. \cite{KKM23}, which is qualitatively similar to the finding
by Gazda \textit{et~al.} \cite{GHF22}. In this article, we show only the results using
N$^4$LO$^+$(550) and N$^4$LO$^+$(400) for the NN interaction, where the number
in parentheses is a cutoff scale in MeV. As for the YN interactions, two versions of the
chiral NLO interactions, NLO13 \cite{NLO13} and NLO19 \cite{NLO19} with a cutoff
scale of 550 MeV parametrized by the J{\"{u}}lich-Bonn group, are employed.
The Nijmegen NSC89 potential \cite{NSC89} is also applied, though it may not be
appropriate to use with the regularized chiral interactions at around 500 MeV.

\begin{figure}[bth]
\includegraphics[width=0.4\textwidth]{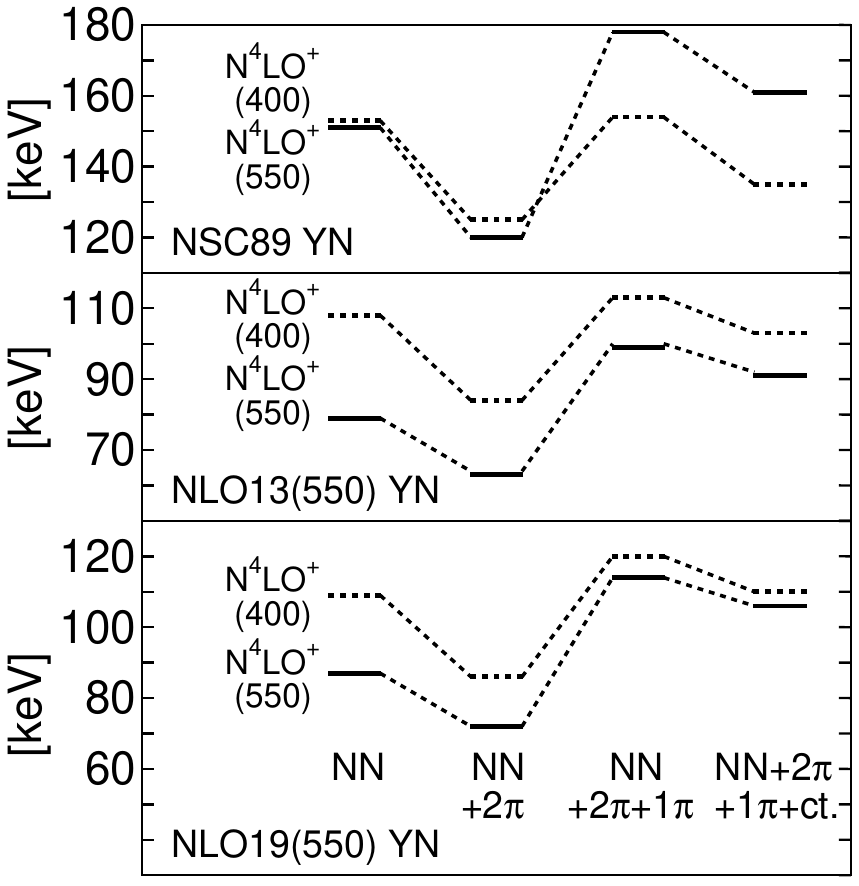}
\caption{Results of the \h3t separation energy obtained by Faddeev calculations.
The NN interactions are the semilocal N$^4$LO$^+$ interaction\cite{RKE18}
with its cutoff scale of 550 and 400 MeV. The lower, middle, and upper panels
show the results for the chiral NLO19, chiral NLO13, and Nijmegen NSC89
YN interactions, respectively. The cutoff scale of the chiral YN interactions is
550 MeV. In each figure, the leftmost entry is the calculation with the NN interactions only.
Other entries are results in which the $2\pi$-exchange, $1\pi$-exchange, and contact
3BFs are added separately.
}
\label{fig:be}
\end{figure}

Calculated results for each YN interaction are depicted in Fig. \ref{fig:be}. The leftmost entry
is the separation energy without 3BFs. The second entry from the left is the result including
the $2\pi$-exchange $\Lambda$NN, which was reported already in Ref. \cite{KKM23}.
The second entry from the right is the result with the $1\pi$-exchange 3BF and
the $2\pi$-exchange $\Lambda$NN. The $1\pi$-exchange 3BF acts attractively due to
the matrix element of the $s$-$d$ pair
of the deuteron wave function, as shown by the folding potential in Sec. 4.
The rightmost entry shows the result in which all the 3BFs are included. The net effect
after including NNLO 3BFs turns out to be sensitive to the NN interaction. In the case of
N$^4$LO$^+$(550) in which the separation is narrower, the 3BFs bring about the attraction
in the order of 20 keV for \h3t. On the other hand, the 3BFs tend to work repulsively
in the case of N$^4$LO$^+$(400). It seems that the 3BFs work attractively in the case
that the \h3t wave function is wide-spreading. With the wave function shrinking,
the net attraction begins to
diminish. The trend of the 3BFs contribution is opposite in the NSC89 YN potential,
for which the interpretation is difficult because the NN interactions are defined in a
low momentum with the cutoff scale of around 500 MeV and the NSC potential
in an entire space.

\section{Summary}
In Ref. \cite{KKM23}, we investigated the effect of the $2\pi$-exchange $\Lambda$NN
3BF in ChEFT in the $_\Lambda^3$H hypernuclei by carrying out Faddeev calculations
that include the 3BF for the first time. In the present article, the remaining $1\pi$-exchange
and contact term $\Lambda$NN 3BFs derived by Petschauer \textit{et~al.} \cite{PET17}
at the NNLO level are addressed. Because the expressions of these 3BFs
in Ref. \cite{PET16} are not readily applicable
in calculating matrix elements of $\Lambda$ hypernuclei, we rewrite them to be applied
in the formula given in Ref. \cite{KKM22}. Then, Faddeev calculations for \h3t are
carried out, including NNLO 3BFs.

The net effect of these 3BFs is found to be small because of the cancellation between the
attractive contribution of the $1\pi$-exchange 3BF and the repulsive ones of the other
3BFs under the tentative assignment for the sign of the coupling constant $D_2'$.
It is interesting to see that the difference in the separation energies with N$^4$LO$^+$(550)
and N$^4$LO$^+$(400) becomes smaller when the 3BFs are included for the chiral YN
interactions.
The attraction from the $1\pi$-exchange 3BF is responsible for the matrix element between
the $s$- and $d$-wave components of the NN wave function due to the tensor part of the
$1\pi$-exchange 3BF. The coupling effect is sizable despite the small $d$-wave
component, as is demonstrated by the folding potential discussed in Sec. 3.  
Naturally, the quantitative results serve only as a reference because of the speculative
nature of the LECs employed.
However, the result provides basic information about the contribution of the 3BFs
because it enables us to infer what changes are induced by modifying each
coupling constant. It is necessary to do similar ab-initio calculations in heavier
$\Lambda$-hypernulcei, although the task is computationally demanding.

\bigskip
{\it Acknowledgements.}
This work is supported by JSPS KAKENHI Grants No. JP19K03849 and No. JP22K03597.

\begin{widetext}
\appendix
\section{Tensor-product  decomposition of $\Lambda NN$ three-body interactions}
 $V_{1\pi,(i,j)}^{(K,\ell_a,\ell_b)}(p,q)$ in Eq. (\ref{eq:1pipw}) for $K=0$, 1 and 2 of the
NNLO $1\pi$-exchange 3BF are given in the following.
($x\equiv \frac{p^2+r_{NN}^2 q^2+m_\pi^2}{2r_{NN}pq}$ with $r_{NN}=\frac{1}{2}$
and $\delta_{\ell_a,\mbox{even}}=\frac{1+(-1)^{\ell_a}}{2}$).
Note that $\Lambda$ hyperon receives a label of 1.
 
\begin{align}
 V_{1\pi,(1,2)}^{K=0,\ell_a,\ell_b}(p,q)=& \delta_{\ell_a\ell_b}(-1)^{\ell_a}\frac{g_A D_1'}{2f_0^2}
  \sqrt{\frac{\hat{\ell_a}}{3}}\left[\delta_{\ell_a 0}-\frac{m_\pi^2}{2r_{NN}pq}Q_{\ell_a}(x)\right],
\\
 V_{1\pi,(1,2)}^{K=1,\ell_a,\ell_b}(p,q)=& 0, \\
 V_{1\pi,(1,2)}^{K=2,\ell_a,\ell_b}(p,q)=& - \frac{g_A D_1'}{2f_0^2}\left[
 \sqrt{\frac{2\hat{\ell_a}\hat{\ell_b}}{15}} \cg0{\ell_a}{\ell_b}{2}
 \left( \frac{p^2}{2r_{NN}pq} Q_{\ell_b}(x)
 +\frac{r_{NN}^2 q^2}{2r_{NN}pq} Q_{\ell_a}(x)\right)\right.  \nonumber \\
 & \left. \hspace{4em}-\sum_{\ell} (-1)^\ell \sqrt{5}
 \hat{\ell}\cg0{\ell}{1}{\ell_a}\cg0{\ell}{1}{\ell_b}
 \sixj{\ell_b}{1}{\ell}{1}{\ell_a}{2} Q_{\ell}(x) \right], \\
 V_{1\pi,(2,3)}^{K=0,\ell_a,\ell_b}(p,q)=& \delta_{\ell_a\ell_b} \delta_{\ell_a,\mbox{even}}
 \frac{2g_A D_2'}{2f_0^2} \sqrt{\frac{\hat{\ell_a}}{3}}
 \left[\delta_{\ell_a 0}-\frac{m_\pi^2}{2r_{NN}pq}Q_{\ell_a}(x)\right], \\
 V_{1\pi,(2,3)}^{K=1,\ell_a,\ell_b}(p,q)=& 0, \\
 V_{1\pi,(2,3)}^{K=2,\ell_a,\ell_b}(p,q)=& -\delta_{\ell_a,\mbox{even}} \frac{2g_A D_2'}{2f_0^2}\left[
 \sqrt{\frac{2\hat{\ell_a}\hat{\ell_b}}{15}} \cg0{\ell_a}{\ell_b}{2}
 \left( \frac{p^2}{2r_{NN}pq} Q_{\ell_b}(x)
 +\frac{r_{NN}^2 q^2}{2r_{NN}pq} Q_{\ell_a}(x)\right)\right.  \nonumber \\
 & \left. \hspace{4em}+\sum_{\ell} \sqrt{5} \hat{\ell}\cg0{\ell}{1}{\ell_a}
 \cg0{\ell}{1}{\ell_b} \sixj{\ell_b}{1}{\ell}{1}{\ell_a}{2} Q_{\ell}(x) \right], \\
 V_{1\pi,(3,1)}^{K=0,\ell_a,\ell_b}(p,q)=& \delta_{\ell_a\ell_b}
 \frac{g_A D_1'}{2f_0^2} \sqrt{\frac{\hat{\ell_a}}{3}}
 \left[\delta_{\ell_a 0}-\frac{m_\pi^2}{2r_{NN}pq}Q_{\ell_a}(x)\right], \\
 V_{1\pi,(3,1)}^{K=1,\ell_a,\ell_b}(p,q)=& 0, \\
 V_{1\pi,(3,1)}^{K=2,\ell_a,\ell_b}(p,q)=& -\frac{g_A D_1'}{2f_0^2}\left[(-1)^{\ell_a}
 \sqrt{\frac{2\hat{\ell_a}\hat{\ell_b}}{15}} \cg0{\ell_a}{\ell_b}{2}
 \left( \frac{p^2}{2r_{NN}pq} Q_{\ell_b}(x)
 +\frac{r_{NN}^2 q^2}{2r_{NN}pq} Q_{\ell_a}(x)\right)\right. \nonumber \\
 & \left. \hspace{4em}+\sum_{\ell} \sqrt{5} \hat{\ell}\cg0{\ell}{1}{\ell_a}
 \cg0{\ell}{1}{\ell_b} \sixj{\ell_b}{1}{\ell}{1}{\ell_a}{2} Q_{\ell}(x) \right].
\end{align}
\end{widetext}

\end{document}